\documentclass[prl,twocolumn,superscriptaddress]{revtex4}

\usepackage{amsmath}
\usepackage{graphicx}
\usepackage{enumerate}

\newcommand{\e}				{\mathrm{e}}
\newcommand{\bx}			{\mathbf{x}}
\newcommand{\bk}			{\mathbf{k}}
\newcommand{\bg}			{\mathbf{g}}
\newcommand{\bG}			{\mathbf{G}}
\newcommand{\bT}			{\mathbf{T}}
\newcommand{\btau}			{\boldsymbol{\tau}}
\newcommand{\matrixZZ}[4]	{\begin{pmatrix} #1 & #2 \\ #3 & #4 \end{pmatrix}}
\newcommand{\sign}			{\mathrm{sgn}}
\newcommand{\perc}			{\%}

\begin{document}

\title
{Edge currents in superconductors with a broken time--reversal symmetry}

\author{Bernd Braunecker}
\affiliation{Department of Physics, Massachusetts Institute of Technology, Cambridge, MA 02139}
\affiliation{Department of Physics, Brown University, Providence, RI 02912}

\author{Patrick A. Lee}
\affiliation{Department of Physics, Massachusetts Institute of Technology, Cambridge, MA 02139}

\author{Ziqiang Wang}
\affiliation{Department of Physics, Boston College, Chestnut Hill, MA 02467}

\date{\today}




\begin{abstract}
We analyze edge currents and edge bands at the surface of a
time--reversal symmetry breaking $d_{x^2-y^2}+id_{xy}$ superconductor.
We show that the currents have large Friedel oscillations with two
interfering frequencies: $\sqrt{2}k_F$ from sub--gap states,
and $2 k_F$ from the continuum. The results
are based independently on a self--consistent slave--boson mean field
theory for the $t-J$ model on a triangular lattice, and on a T--matrix
scattering theory calculation. The shape of the edge--state band,
as well as the particular frequency $\sqrt{2}k_F$ of the Friedel
oscillations are attributes unique for the $d_{x^2-y^2}+id_{xy}$
case, and may be used as a fingerprint for its identification.
Extensions to different time--reversal symmetry breaking
superconductors can be achieved within the same approach.
\end{abstract}


\maketitle


Superconductors which break time-reversal symmetry have attracted a 
great deal of attention recently.  The most prominent example is 
Sr$_2$RuO$_4$, which has a $p$-type complex order parameter \cite{Mackenzie03}. 
There have been theoretical suggestions that 
Na$_x$CoO$_2 \cdot y$H$_2$0 belongs to this class, but with complex 
$d$-wave symmetry \cite{Baskaran03,Wang03,Kumar03}.  
It is known that interesting edge effects are induced by the 
surface in such superconductors. 
First, the surface induces the 
appearance of an edge band inside the
superconducting gap. Second, 
a chiral edge current parallel to the
surface appears.
As shown in several independent semi--classical calculations
\cite{Honerkamp98,Senthil99,Matsumoto99,Horovitz03},
the shape of the dispersion $\epsilon_b$ of the edge states
depends on the symmetry of the superconductor and may be used 
as a fingerprint for the latter.
In this letter, we quantitatively address this problem and point out 
a novel interference in the edge current which distinguishes between
$d_{x^2-y^2}+i d_{xy} (\equiv d+id')$ 
and complex $p$ time--reversal symmetry breaking superconductors.
We perform a self--consistent solution of the
$t-J$ model on a triangular lattice which has a
$d+id'$ superconducting phase as the solution of its
slave--boson mean-field theory \cite{Baskaran03,Wang03,Kumar03}.
In order to explain the novel results uncovered by the 
numerical solution, we analytically solve a quantum continuum model
by treating the quasiparticle scattering on the surface using  
an extension of the T--matrix formulation of \cite{WangWang04}.

\begin{figure}[p]
\begin{center}
	\includegraphics[width=0.8\columnwidth]{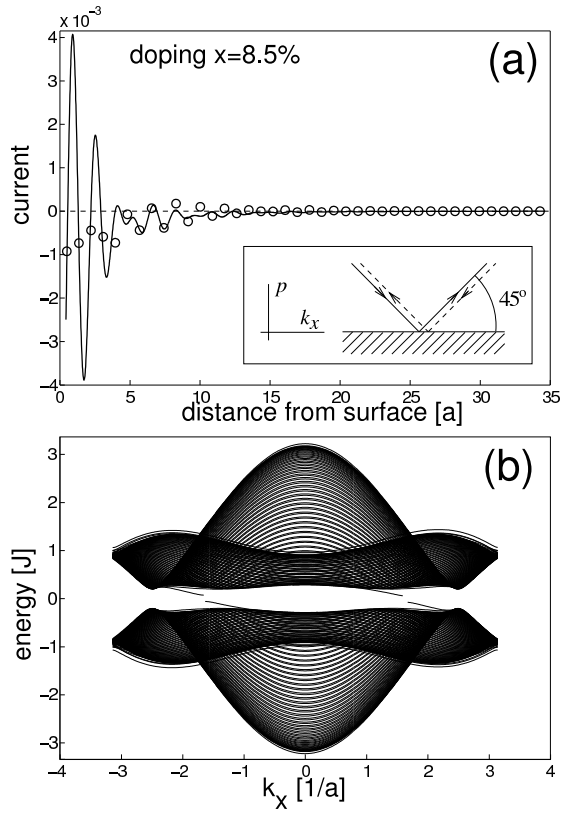}
	\caption{(a) Current (in dimensionless units) from the
	self--consistent calculation
	(circles) and the prediction of Eqs.~\eqref{eq:j_sg} and
	\eqref{eq:j_cont} multiplied by an exponential decay
	factor $\exp(-y/\xi_\text{cur})$ (full line).
	The inset shows the quasiclassical Andreev--scattering on the
	surface leading to the $\sqrt{2}k_F$ oscillations (full lines:
	particle scattering, dashed lines: hole scattering).
	(b) Spectrum $E_{n,k_x}$ as a function of the momentum $k_x$;
	$n$ is the quantization along the $y$--direction. Continuum
	states form the hat--like structures, the sub--gap states have
	roughly parabolic forms. The small gap at zero energy is a
	finite size effect due to the coupling between the two edges.
	\label{fig:current}}
\end{center}
\end{figure}

The results are summarized as follows (see Fig.~\ref{fig:current}):
The surface induces edge currents and an edge
band in agreement with the semi--classical
prediction of two parts of a parabola
\cite{Horovitz03},
$\epsilon_b(k_x) = - \Delta \sign(k_x) (2 k_x^2/k_F^2 - 1)$.
As a surprising novel result, the size and direction of the current
changes rapidly with distance from the surface.
This is interpreted as Friedel oscillations of
\emph{two} frequencies, $2 k_F$ and $\sqrt{2} k_F$.
The $2 k_F$ oscillations are the usual Friedel oscillations
of the continuum states. The $\sqrt{2} k_F$ oscillations
are due to the zero energy mode of the sub--gap band, $\epsilon_b = 0$,
and correspond to a quasiparticle scattering on the surface at an incident
angle of $45^\circ$ \cite{Senthil99} (Fig.~\ref{fig:current} (a) inset).
Most notably, the amplitudes of both oscillations are identical,
and comparable to the nonoscillating part of the current; from
the self--consistent numerical results we see that they are indeed
large enough to reverse the current direction in some regions
close to the surface. The overall
magnitude of the edge current, therefore, is considerably smaller than
could be expected naively. Our calculations lead to an integral
edge current $I$ of the order of 60nA, which also corresponds
roughly to its maximal amplitude. Such a current may be captured
by measuring the induced magnetic field $B$.
For our model calculations, we estimate its maximum at the surface as
$B = \mu_0 I/ 2\pi \xi \sim 0.1\mathrm{G}$,
where $\xi$ is the coherence length of the superconductor.


The $t-J$ model on a two--dimensional triangular lattice
is an example of a microscopic
model with a $d+id'$ superconducting phase as the result of the
slave--boson mean--field theory \cite{Baskaran03,Wang03,Kumar03}.
We solve the mean--field theory self--consistently in real space
following the approach by \cite{WangZ02} in the presence of
two surfaces on the top and bottom of the system.
The model Hamiltonian is the $t-J$ model plus a long--range Coulomb repulsion,
\begin{equation}
	H =
	- t \sum_{\langle i j\rangle}
	  ( c_{i\sigma}^\dagger c_{j\sigma} + \text{h.c.})
	+ J \sum_{\langle i j\rangle}
	  (\mathbf{S}_i \cdot \mathbf{S}_j - \frac{n_i n_j}{4})
	+ \sum_i V_i n_i,
\end{equation}
where $\langle i j \rangle$ runs over the nearest neighbors lattice sites
of an equilateral triangular lattice, $c_{i\sigma}$ are the electron
operators, and $\mathbf{S}_i$ the spin $1/2$ operator. The model is
completed by the constraint of no double occupancy per site,
$n_i = \sum_\sigma c_{i\sigma}^\dagger c_{i\sigma} \le 1$.
$J>0$ is the antiferromagnetic exchange interaction, and $t$ the hopping
integral.
Because of the missing particle--hole symmetry on the triangular
lattice, the sign of $t$ is important \cite{Baskaran03,Wang03}.
We choose $t/J = -3$, corresponding to electron doping
of the system.
$V_i = V_c \sum_{j \neq i}(n_j - \bar{n})/|\mathbf{r}_i-\mathbf{r}_j|$,
is the Coulomb potential felt by the particle at the site $i$,
where $\bar{n}$ is the average density, $j$ runs over all lattice
sites, and $V_c \approx 5 J$ \cite{WangZ02}.
The long--range Coulomb interaction is necessary to overrule the
inherent tendency of the mean--field theory to phase separate.
In a uniform system, $V_i$ vanishes.
On the lightly doped triangular lattice, however, the Coulomb
interaction alone is not strong enough against the
phase separating instabilities. The key insight for a
stabilization is the correction of
the mean--field theory by a Jastrow--like modification of the
hopping parameter $t \to t \exp(- w | \sqrt{(1-n_i)(1-n_j)} - (1-\bar{n})|)$
for nearest neighbor sites with $w \sim 5$. This factor suppresses
the gain in kinetic energy by clustering carriers on
neighboring sites. For small fluctuations in the carrier concentrations,
it is close to 1. Its precise form otherwise has not much influence.

In the slave--boson formulation, the electron operator is decomposed
as $c_{i \sigma} = f_{i\sigma} b_i^\dagger$, where $f_{i\sigma}$
is a spin carrying fermion, and $b_i$ a charge carrying boson operator.
The constraint $n_i \le 1$ becomes the identity
$\sum_\sigma f_{i\sigma}^\dagger f_{i\sigma} + b_i^\dagger b_i = 1$,
which must be fulfilled at each site $i$, and which can be included in the
action with Lagrange multipliers $\lambda_i$.

In the superconducting phase, the bosons $b_i$ are
condensed and directly related to the local carrier concentrations,
$x_i = (1-n_i) = b_i^2$.
The (Jastrow--corrected) mean--field theory can be derived by a variational
Ansatz \cite{Brinckmann01}, and
leads to a Hamiltonian
for the fermion fields $f_{i\sigma}$, expressed in terms of the $x_i$,
the hopping parameter
$\chi_{ij} = \langle f_{i\sigma}^\dagger f_{j\sigma}\rangle$,
and the pairing parameter
$\Delta_{ij} = \langle f_{i\uparrow} f_{j\downarrow} -
f_{i\downarrow} f_{j\uparrow}\rangle$.
The theory is solved self--consistently by iteration:
For random initial order parameters, we solve
the Bogoliubov--de~Gennes equations by diagonalization
of the Hamiltonian. The resulting spectrum and wave functions
are used to reconstruct the order parameters and the local densities.
The Lagrange multipliers $\lambda_i$ are calculated explicitly
by minimizing the action with respect to the condensed bosons $b_i$.
This is repeated until convergence and requires for the triangular
lattice at low dopings ($< 15\perc$) about 500 to 1000 iterations.

We study lattices consisting in a strip with periodic boundary
conditions along the $x$--direction and, to model the two surfaces,
open boundary conditions along the $y$--direction.
To reduce the interference between the edges, we choose systems in
which the surfaces are far apart (typically $\sim 100$ lattice sites).
Finite size constraints are important and distort the perfect
$d+id'$--symmetry in that the phase difference of $\Delta_{ij}$
between lattice directions deviates from $2 \pi/3$, or that the ratio
$\mathrm{Re}\Delta_{ij}/\mathrm{Im}\Delta_{ij}$ in the bulk
deviates from unity.
To minimize these effects, we choose a sufficiently large extension
along the $x$--direction of $N_x = 24$. The computational expense
then limits the $y$--extension to about 100 sites.

In Fig.~\ref{fig:current} (a) (circles) we show the resulting
edge current.
It has oscillations with a large amplitude which lead at some
layers to the reversal of the current direction. The spatial period
of the oscillations is roughly twice the lattice spacing, but
beatings indicate a frequency mixture. The multiplication of the
displayed current by $t e/\hbar$ ($e =$ electron charge) provides
the current in Amp\`{e}res. With $t=-3J$ and $J=20\textrm{meV}$
(as estimated for the \mbox{Na$_x$CoO$_2$} superconductor, which
was considered as a candidate for the $d+id'$ symmetry
\cite{Baskaran03,Wang03,Kumar03}), the overall current, integrated along the
direction perpendicular to the surface, is about 60nA (setting
an order of magnitude), and is concentrated
within a few times the decay length of $\xi_\text{cur} \approx 5 a$
(with $a \approx 3\mathrm{\AA}$ the lattice constant for the equilateral
triangular lattice).

Part (b) of the figure shows the spectrum of the system
as a function of $k_x$ after a Fourier transformation of the
real--space result. We have suppressed the edge states at one of the
two surfaces by inspection of the support of the wave functions.
An asymmetric sub--gap band is clearly visible.
Its shape deviates slightly from the parabola predicted from the
semi--classical arguments as an effect of the discrete and finite lattice
and the variation of the magnitude of $\Delta$ close to the edge.
In particular, a small gap persists as the result of the limited
system size and the interference between the edges. The gap
closes with increasing distance of the surfaces.

A similar computation can be performed for a lattice with a single
site defect \cite{BB04b}. Instead of a band, a pair of bound states
appears at the defect site. For similar interaction parameters, the
edge current is concentrated on the immediate neighbors of the defect
site only, and has a value of about 100nA. This leads to a magnetic
field of the order of 1G on the defect site.


The large amplitude of the oscillations seems to be surprising, but
can be derived within a continuum model for the scattering on
the surface. We show that the oscillations are Friedel oscillations
induced by the scattering on the surface. The calculation must go
beyond the semi--classical arguments, and we follow the T--matrix
calculation by \cite{WangWang04}, which we extend for the calculation
of the current density $j(y)$.
Even though we focus on the $d+id'$ symmetry, the approach is general
and can be extended to any given symmetry of the superconductor.

We consider a semi--infinite two--dimensional $d+id'$ superconductor
in the $(x,y)$--plane with its surface at $y=0$, and the superconductor
at $y>0$.
The current can be obtained from the causal (time--ordered) Green's function,
$G^c(\bx,\bx';t)$, by the expression
\begin{equation} \label{eq:def_j}
	j(\bx) = \frac{1}{2m} \sum_\sigma (\nabla_\bx - \nabla_{\bx'})
	G^c_\sigma(\bx,\bx';t=0^-)\big|_{\bx'=\bx},
\end{equation}
with $\sigma$ the spin projections, and $m$ the mass of the carriers.
In contrast to \cite{WangWang04}, causal instead of retarded Green's
functions must be used.

For the $d+id'$ superconductor, we write the gap function in the
simplified, position independent form $\Delta_\bk = \Delta \e^{2i \theta}$,
where $\theta$ is the angle of the momentum $\bk$ to a given axis.
Since we intend here to provide an explanation to the observed
oscillations while keeping the calculation as simple as possible,
we assume $|\Delta|$ to be constant even in the vicinity of the
surface. The variability of $\Delta$ is taken into account, however, in the
self--consistent numerical solution of the Bogoliubov--de~Gennes equations.
Since the amplitude of the gap is independent of the direction,
the orientation of the surface is of no importance.
In this approximation, the Fermi surface is assumed to be cylindrical.
The surface is modeled by a line of scattering centers with
elastic potentials $V$, which eventually we let tend to infinity.
The preserved translation symmetry along the $x$--axis allows us to keep
the momenta parallel to the surface, $k_x$, so that the scattering
equation reads
\begin{equation}
	\bG^c_{k_x}(y,y';\omega)
	= \bg^c_{k_x}(y-y';\omega)
	+ \bg^c_{k_x}(y;\omega) \bT_{k_x}(\omega)\bg^c_{k_x}(-y';\omega),
\end{equation}
where the bold symbols denote functions in the particle--hole Nambu
space, and $\omega$ is the frequency.
The free causal Green's function, $\bg^c_{k_x}(y)$, is
\begin{multline}  \label{eq:bgc}
	\bg^c_{k_x}(y;\omega)
	=
	- \frac{\pi N_0}{p \sqrt{\Delta^2-\omega^2}-i0}
	\e^{-\frac{y k_F}{4p \xi}\sqrt{1-(\omega/\Delta)^2}}
	\\
	\times
	\bigl\{
		\omega\cos(py)\btau_0
		+ \Delta \sum_{\nu=\pm} \cos(py + \nu 2 \theta) \btau_\nu
	\bigr\},
\end{multline}
where $N_0$ is the density
of states at the Fermi energy, $k_F$ the Fermi vector,
$p = \sqrt{k_F^2-k_x^2}$, and $\xi$ the coherence length.
Here $k_x = k_F \cos(\theta), p = k_F \sin(\theta)$,
$\theta \in [0,\pi]$.
The $\btau_\pm = \btau_x \pm i \btau_y$ are the Pauli matrices
in the Nambu space; $\btau_0$ is the unit matrix.

The T--matrix is defined by
$\bT^{-1}_{k_x}(\omega) = V \btau_z - \bg^c_{k_x}(\omega)$.
To model the surface, we let $V\to \infty$. With
$\Omega = \sqrt{\Delta^2-\omega^2}-i 0$ and Eq.~\eqref{eq:bgc}
this leads to
\begin{equation} \label{eq:T1}
	\bT =
	\frac{p \Omega/\pi N_0}{\omega^2-\Delta^2 \cos^2(2\theta)(1-i0)}
	\matrixZZ{\omega}{-\Delta\cos(2\theta)}{-\Delta\cos(2\theta)}{\omega}.
\end{equation}
In this expression we still have scattering between physical $y>0$ states and
unphysical $y<0$ states. It contains two sub--gap bands with the energies
$\epsilon_b^{(1,2)}(k_x) = \pm \Delta \cos(2\theta)$
for the states defined for $y>0$ and $y<0$ respectively. The T--matrix must
be split into two parts acting on those states separately.
For the time--reversal symmetry breaking superconductor, we expect an
asymmetry of these bands with respect to the sign of $k_x$, leading to
the only possible choice
\begin{equation}
	\epsilon_b^{(1)}(k_x)
	= - \sign(k_x) \Delta \cos(2\theta)
	= - \sign(k_x) \Delta (2 \tfrac{k_x^2}{k_F^2}-1),
\end{equation}
and $\epsilon_b^{(2)} = - \epsilon_b^{(1)}$.
This is precisely the expression obtained from the semi--classical
models \cite{Honerkamp98,Horovitz03}, and we can focus on the
sub--gap band $\epsilon_b^{(1)}(k_x) \equiv \epsilon_b(k_x)$ only,
relevant for $y>0$.
The splitting is achieved by writing
$1/[(\omega-\epsilon_b^{(1)})(\omega-\epsilon_b^{(2)})]$ as the
sum and difference of $1/(\omega-\epsilon_b^{(1,2)})$, such that
the remaining factors in the T--matrix \eqref{eq:T1} compensate any
unphysical singularity appearing from this decomposition.
The T--matrix becomes, disregarding the $y<0$ part,
\begin{equation}
	\bT_{k_x}(\omega)
	=
	\frac{p \Omega/2 \pi N_0}{\omega-\epsilon_b(k_x)(1-i0)}
	\matrixZZ{1}{\sign(k_x)}{\sign(k_x)}{1}.
\end{equation}
In $(k_x,y,\omega)$-space, the current density \eqref{eq:def_j} then reads
(using $N_0=m/2\pi$)
\begin{multline} \label{eq:j_kxyomega}
	j(k_x,y;\omega)
	= \frac{2i k_x}{m} 
[\bg^c_{k_x}(y;\omega)\bT_{k_x}(\omega)\bg^c_{k_x}(-y;\omega)]_{11}
	\\
	= \frac{i k_x}{2 p \Omega}
	  \frac{\e^{-\frac{y k_F}{2 p \xi}\sqrt{1-(\omega/\Delta)^2}}}{\omega-\epsilon_b(1-i0)}
	  \Bigl\{
	  	\cos^2(p y) (\omega-\epsilon_b)^2
		\\
		- \Delta \sin(2py) \left|\sin(2\theta)\right| 
(\omega-\epsilon_b)
		+ \Delta^2 \sin^2(py)\sin^2(2\theta)
	  \Bigr\}.
\end{multline}
The $\omega$--integration leads to two contributions, the
sub--gap poles, which exist for $\omega=\epsilon_b<0$ only,
and an integral running along the continuum state branch cut
at $\omega<-\Delta$.
For the sub--gap states, we obtain
(see also \cite{Horovitz03}):
\begin{multline}
	j^\text{sg}(y) = -\frac{\Delta}{4 \pi} \int_{-k_F}^{k_F} \mathrm{d}k_x
	\frac{k_x}{p} \e^{-\frac{y}{\xi}|\cos(\theta)|}
	\sin^2(py)
	\\
	\times
	|\sin(2\theta)| \Theta(-\epsilon_b),
\end{multline}
with $\Theta$ the unit step function. Close to the surface, $y\ll\xi$,
we can replace the exponential by unity. At $y\approx 0$, the
current grows as $y^2$. For $y k_F \gg 1$ but still
$y \ll \xi$, the integrand oscillates rapidly, and, to leading
order in $1/k_F y$, $j^\text{sg}$ can be estimated as
\begin{equation} \label{eq:j_sg}
	j^\text{sg}(y)
	= \frac{\Delta k_F}{4 \pi}
	\Bigl[ \frac{1}{3\sqrt{2}}-\frac{\sin(\sqrt{2} k_F y)}{2 k_F y}\Bigr],
\end{equation}
i.e. the spatial oscillations are determined by the zero energy mode of
quasiparticle scattering on the surface, for which $p=k_F/\sqrt{2}$
\cite{Senthil99} (see Fig.~\ref{fig:current} (a), inset).
Constant contributions at the artificial
cutoffs at $k_x = \pm k_F$ have been neglected.
For $y$ exceeding the coherence length $\xi$, this current is
suppressed exponentially. This allows us to identify $\xi_\text{cur} = \xi$.

In a time--reversal symmetry breaking superconductor the continuum states carry
a current, too.
Their contribution is given from the $\omega$--integral running along 
the branch
cut at $(-\infty,-\Delta)$. We restrict the integration over $k_x$ to
positive values, and keep only terms which are even in $k_x$.
This yields ($\omega \to -\omega$)
\begin{multline} \label{eq:j_cont0}
	j^{\text{c}}(y)
	=
	\frac{N_0}{\pi m}
  	\int_0^{k_F}
	\frac{\mathrm{d}k_x k_x \epsilon_b}{p}
	\int_{\Delta}^{\infty} \mathrm{d}\omega
	\frac{\cos\bigl(\tfrac{y k_F}{2p \xi}
	      \sqrt{(\frac{\omega}{\Delta})^2-1}\bigr)}%
	     {\sqrt{\omega^2-\Delta^2}}
	\\
	\times
	\frac{(\omega^2-\Delta^2) \cos^2(py)
		  + \Delta^2\sin^2(2\theta)\cos(2py)}%
	     {(\omega^2 - \Delta^2) + \Delta^2 \sin^2(2\theta)}.
\end{multline}
The integral weight is dominated by the singularity
$1/\sqrt{\omega^2-\Delta^2}$ and is concentrated at the physically
relevant values of $\omega \approx \Delta$. In this region, the
last factor in Eq.~\eqref{eq:j_cont0} varies slowly about $\cos(2py)$, and
we replace it by this quantity.
For $y\gg \xi$, the $\omega$--integration provides a Bessel function
which accounts for the required exponential decay. At $y=0$, however,
the $\omega$--integration reduces to the integration of
$1/\sqrt{\omega^2-\Delta^2}$, which is divergent.
This is an artifact of the approximation of
a constant $N_0$. The integral must be cut off at the band--edge
where the density of states vanishes and, therefore, is
given solely by its value at $\omega=\Delta$,
$\int\mathrm{d}\omega/\sqrt{\omega^2-\Delta^2}|_{\Delta} = \pi/2$.
For $y$ close to the surface, we replace the $\omega$--integral
by this value. The remaining $k_x$ integration is
elementary. We obtain for $y \ll \xi$, to leading order in $1/k_Fy$,
\begin{equation} \label{eq:j_cont}
	j^\text{c}(y) =
	\frac{\Delta k_F}{4 \pi} \frac{\sin(2 k_F y)}{2 k_F y}.
\end{equation}
The full current density, $j(y)$, is the sum of Eqs.~\eqref{eq:j_sg} and
\eqref{eq:j_cont}, and shows Friedel--like spatial oscillations with the
two frequencies $\sqrt{2} k_F$ and $2 k_F$.

Even though details vary in real physical systems and, as shown above,
in fully self--consistent solutions, these expressions make the following
precise statements:
\emph{i)} The edge state dispersion is identical to the semi--classical
results \cite{Honerkamp98,Horovitz03}.
\emph{ii}) There are two interfering Friedel oscillations of the same
amplitude with the frequencies $\sqrt{2}k_F$ and $2 k_F$, where
the former is determined by the condition $\epsilon_b(k_x) = 0$.
\emph{iii}) Most notably, the amplitude of the oscillations is comparable
to the non--oscillating term in the current, i.e. it can reverse
the direction of the current in some regions close to the surface.
The exact ratio of the amplitudes cannot be captured with the
present (not self--consistent) calculation.
Yet the self--consistent numerical calculation above shows
that this is indeed the case (Fig.~\ref{fig:current}).
\emph{iv}) The current amplitude is proportional to $\Delta$ and $k_F$.

Note that Friedel oscillations with the same two frequencies
are found for the carrier density, $n(y)$, since it connects to the
Green's function through $n(y) = -2 i [\bG^c(y,y;t=0^-)]_{11}$

We also remark that in the spin--zero sub--gap band of a
$p_x + ip_y$ superconductor with $\Delta_k = \Delta \e^{i \theta}$
\cite{WangWang04} the same calculation leads to $\epsilon_b \sim k_x$,
which vanishes at $p=k_F$.
This provides a frequency $2 k_F$ for the Friedel oscillations from
the sub--gap contribution and is indistinguishable from the continuum
state oscillations.

 From the spectrum in Fig.~\ref{fig:current} (b), we obtain
$k_x \approx 2.2/ a$ at $\epsilon_b = \Delta$,
and $k_x \approx 1.6/ a$ at $\epsilon_b = 0$, with a ratio of
precisely $\sqrt{2}$. With these two values,
and $\Delta$ taken from the bulk of the lattice, we can superpose the
numerical result with the theoretical expressions \eqref{eq:j_sg}
and \eqref{eq:j_cont}. We complete the theoretical expression by an
exponential decay factor $\e^{-y/\xi_\text{cur}}$.
While the the amplitude and frequencies are fixed by $\Delta$ and $k_F$,
we fit the data for $\xi_\text{cur}$, the effective distance between
the surface and the first lattice layer, and the value of the nonoscillating
part (replacing the $1/3\sqrt{2}$). The resulting curve shows a nice
agreement between the theoretical and numerical currents
(Fig. \ref{fig:current} (a)). We deduce a $\xi_\text{cur} \approx 5 a$,
and a nonoscillating part, which is an order of magnitude below the
prediction of $1/3\sqrt{2}$.
The integration of both the theoretical and numerical currents leads
to an integral current of $60\mathrm{nA}$.

To conclude, we have shown with two independent calculations
that two types of Friedel oscillations exist in the time--reversal
symmetry breaking superconductor, and that they play an essential
role in the surface effects. The expected integral edge current and the
induced magnetic field is reduced by about an order of magnitude
with respect to the estimates for a nonoscillating prediction,
which has important consequences for experiments.


We thank Sung--Sik Lee, J. B. Marston, and K. A. Moler
for helpful discussions.
BB acknowledges the support of the Swiss National Science Foundation,
PAL acknowledges the support of the DOE grant No. DE-FG02-03ER45076,
ZW acknowledges DOE grant No. DE-FG02-99ER45747 and ACS
grant No. 39498-AC5M.



\end{document}